\documentclass[reprint,amssymb, amsmath, aps, superscriptaddress, showpacs, footinbib,  prb]{revtex4-1}
\pdfoutput=1
\usepackage{graphicx}

\newcommand{\iso}{\text{iso}}

\newcommand{\Cels}{$^{\circ}$C}

\begin{document}

\title{Spiral ground state against ferroelectricity in the frustrated magnet BiMnFe$_2$O$_6$}

\author{Artem M. Abakumov}
\email{Artem.Abakumov@ua.ac.be}
\affiliation{EMAT, University of Antwerp, Groenenborgerlaan 171, B-2020 Antwerp, Belgium}

\author{Alexander A. Tsirlin}
\email{altsirlin@gmail.com}
\affiliation{Max Planck Institute for Chemical Physics of Solids, N\"{o}thnitzer
Str. 40, 01187 Dresden, Germany}

\author{Juan Manuel Perez-Mato}
\affiliation{Dpto de Fisica de la Materia Condensada, Facultad de Ciencia y Tecnologia, Universidad del Pais Vasco, Apdo 644, Bilbao 48080, Spain}

\author{Vaclav Pet{\v r}i{\v c}ek}
\affiliation{Institute of Physics, ASCR, v.v.i, Na Slovance 2, 182 21 Praha 8, Czech Republic}

\author{Helge Rosner}
\affiliation{Max Planck Institute for Chemical Physics of Solids, N\"{o}thnitzer
Str. 40, 01187 Dresden, Germany}

\author{Tao Yang}
\author{Martha Greenblatt}
\affiliation{Department of Chemistry and Chemical Biology, Rutgers, the State University of New Jersey, 610 Taylor Road Piscataway, NJ 08854-8087}


\begin{abstract}
The spiral magnetic structure and underlying spin lattice of BiMnFe$_2$O$_6$ are investigated by low-temperature neutron powder diffraction and density functional theory band structure calculations. In spite of the random distribution of the Mn$^{3+}$ and Fe$^{3+}$ cations, this centrosymmetric compound undergoes a transition into an incommensurate antiferromagnetically ordered state below $T_N\simeq 220$~K. The magnetic structure is characterized by the propagation vector $\mathbf k=[0,\beta,0]$ with $\beta\simeq 0.14$ and the $P22_12_11'(0\beta0)0s0s$ magnetic superspace symmetry. It comprises antiferromagnetic helixes propagating along the $b$-axis. The magnetic moments lie in the $ac$ plane and rotate about $\pi(1+\beta)\simeq 204.8$~deg angle between the adjacent magnetic atoms along $b$. The spiral magnetic structure arises from the peculiar frustrated arrangement of exchange couplings in the $ab$ plane. The antiferromagnetic coupling along the $c$-axis cancels the possible electric polarization, and prevents ferroelectricity in BiMnFe$_2$O$_6$. 
\end{abstract}

\maketitle

\section{Introduction}
\label{introduction}
The coupling between magnetism and ferroelectricity is one of the intriguing phenomena in solid-state physics.\cite{wang2009} Apart from the ongoing studies of the underlying microscopic mechanisms,\cite{katsura2005,moskvin2008} the effect itself is relevant for applications,\cite{wang2009} and stimulates extensive experimental work on diverse systems varying from bulk transition-metal compounds \cite{kimura2003,ikeda2005,kagawa2010} to heterostructures.\cite{wu2010} Magnetoelectric effects in bulk systems typically conform to one of the two following scenarios: (i) the magnetism arises from transition-metal cations with a partially filled $d$ shell, and the ferroelectricity is driven by lone-pair cations, such as Bi$^{3+}$ or Se$^{4+}$ (Refs.~\onlinecite{khomskii2006,hill2002}), or (ii) electronic effects [a spiral (helicoidal) magnetic structure or a charge ordering] break the symmetry and cause ferroelectricity.\cite{khomskii2006,cheong2007,brink2008} The former mechanism ensures large electric polarization, which, however, is weakly coupled to the magnetism. The latter scenario provides a strong coupling, but a small electric polarization. The combination of the two approaches is clearly advantageous, but difficult to achieve. The best known example is BiFeO$_3$,\cite{catalan2009} which exhibits a plethora of interesting effects related to the coupling between ferroelectricity and magnetism.\cite{zhao2006,rovillain2010} 

\begin{figure}
\includegraphics{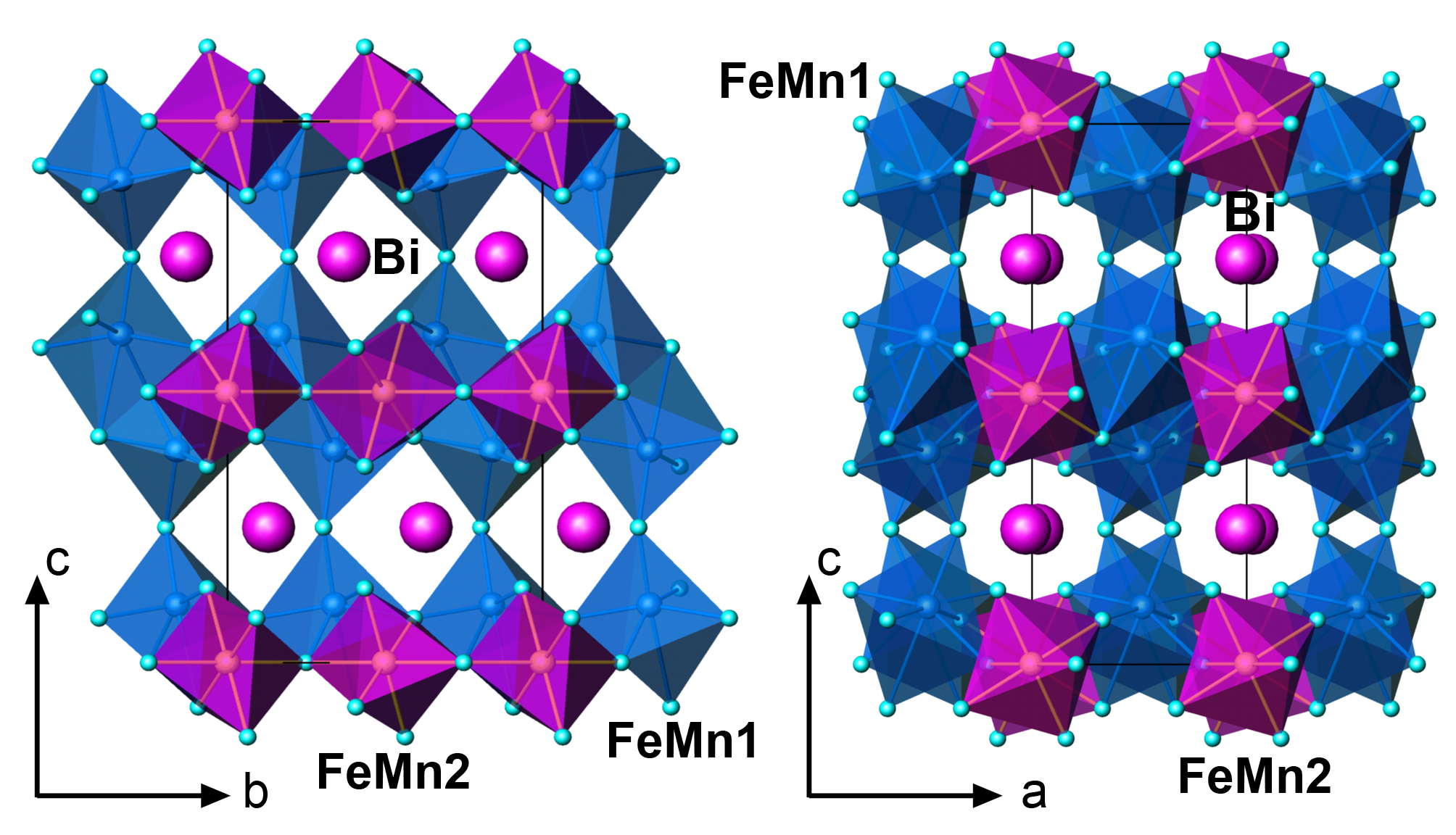}
\caption{\label{structure}
(Color online) The crystal structure of BiMnFe$_2$O$_6$. The Fe and Mn cations are situated in the oxygen octahedra (in the color version, blue and violet polyhedra denote the FeMn1 and FeMn2 positions, respectively).
}
\end{figure}
The recently discovered complex oxide BiMnFe$_2$O$_6$ represents another system combining the two scenarios plausible for multiferroicity: the lone-pair Bi$^{3+}$ cation and the spiral magnetic ground state. Neither of the two, however, lead to ferroelectricity. The polar displacements of Bi$^{3+}$ are ordered in an antiferroelectric manner,\cite{yang2010} whereas the spiral magnetic structure is non-polar due to a strong antiferromagnetic coupling along the crystallographic $c$ direction.

The unique crystal structure of BiMnFe$_2$O$_6$ features fragments of the hypothetical $\textit{hcp}$ oxygen-based MO building blocks that are related by a mirror operation into a polysynthetically twinned structure.\cite{yang2010} The Fe$^{3+}$ ($d^{5}$) and Mn$^{3+}$ ($d^{4}$) cations are nearly randomly distributed over two crystallographically distinct positions, both octahedrally coordinated by oxygen atoms (Fig.~\ref{structure}). The octahedra are interconnected into a framework through corner-, edge-, and face-sharing, providing complex paths for the magnetic exchange. In spite of the random distribution of the Fe and Mn cations, a long-range magnetic order was reported in BiMnFe$_2$O$_6$ below $T_N\simeq 220$~K, according to $^{57}$Fe M\"ossbauer spectroscopy, magnetic susceptibility, and heat capacity measurements. The low-temperature neutron powder diffraction revealed the incommensurate propagation vector of the magnetic structure, yet no details of the ground-state spin arrangement have been reported.\cite{yang2010}
	
Spiral magnetic structures are capable of inducing ferroelectricity in a number of transition-metal oxides, such as TbMnO$_3$,\cite{kenzelmann2005} Ni$_3$V$_2$O$_8$,\cite{lawes2004,lawes2005} MnWO$_4$,\cite{heyer2006} and Ba$_{0.5}$Sr$_{0.5}$Zn$_2$Fe$_{12}$O$_{22}$.\cite{kimura2005} In contrast to the aforementioned compounds, BiMnFe$_2$O$_6$ remains paraelectric below $T_N$.\cite{yang2010} To understand the lack of ferroelectricity, we studied the magnetic structure and explored the underlying frustrated spin lattice. Although the strong frustration leads to the formation of magnetic helices propagating along $b$, the unfrustrated antiferromagnetic (AFM) coupling along $c$ induces the 222 point symmetry of the magnetically ordered state and cancels the possible electric polarization.

\section{Methods}
\label{methods}
The powder sample of BiMnFe$_2$O$_6$ was synthesized by high-temperature solid-state reaction in air. Stoichiometric amounts of raw materials (Bi$_2$O$_3$, Mn$_2$O$_3$, Fe$_2$O$_3$) were ground thoroughly and heated up to 800~\Cels\ in 5~h. After annealing for 10~h, the powder was reground, pressed into a pellet, and heated at 1000~\Cels\ for 100~h with several intermediate grindings.
\begin{figure}
\includegraphics{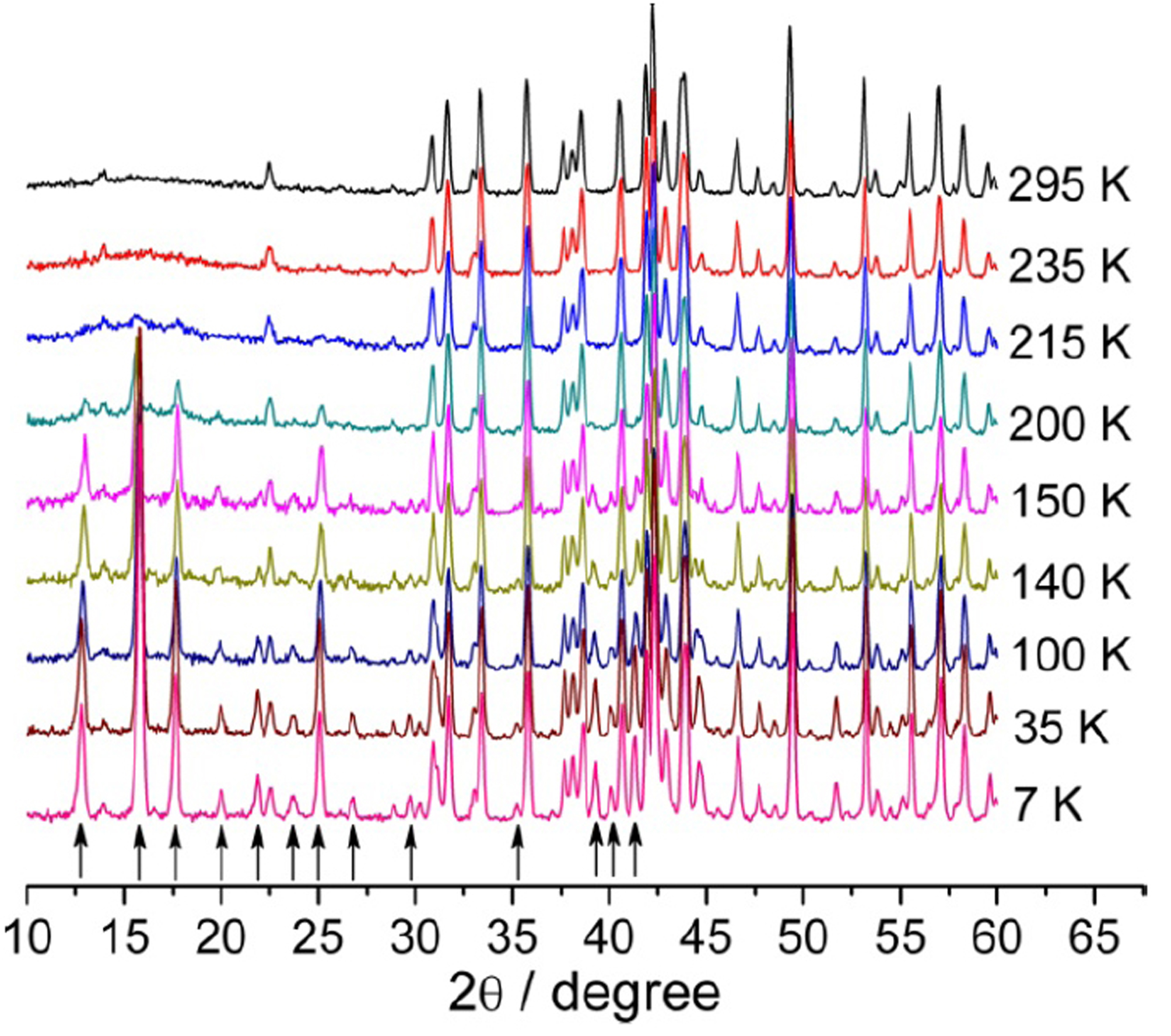}
\caption{\label{NPD}
(Color online) Neutron diffraction patterns for BiMnFe$_2$O$_6$ at different temperatures. The arrows show the magnetic reflections.
}
\end{figure}

Neutron powder diffraction (NPD) data were collected on a $\,\simeq\!12$~g sample of BiMnFe$_2$O$_6$, contained in a 9.5~mm diameter vanadium can. A closed cycle He refrigerator was used for temperature control. Patterns were collected with the BT-1 32-detector high-resolution neutron powder diffractometer at the National Institute of Standards and Technology Center for Neutron Research, Gaithersburg, MD. A Cu(311) monochromator with a $90^{\circ}$ takeoff angle and 15 min in-pile collimation was used. The neutron wavelength was 1.5402(1)~\r A. Data from the 32 detectors were combined to give pseudo one-detector data over a total scan range of $3^{\circ}\leq 2\theta\leq 167.75^{\circ}$ with a step size of $0.05^{\circ}$ ($2\theta$). The magnetic structure was analyzed with the JANA2006 program.\cite{jana2006} 

Density functional theory (DFT) band structure calculations were performed in the full-potential local-orbital (\texttt{FPLO}) code.\cite{fplo} We used local density approximation (LDA)\cite{pw92} supplied with a mean-field (DFT+$U$) correction for correlation effects in the Fe/Mn 3$d$ shell. The $k$-mesh comprised 192 points for the 40-atom unit cell and 64 points for the 80-atom supercell. The correlated shell was parameterized with an effective on-site Coulomb repulsion $U_d=5$~eV and exchange $J_d=1$~eV,\cite{leonov2004,abakumov2010} whereas the double-counting was corrected in the fully-localized-limit (atomic limit) fashion. To evaluate individual exchange couplings, total energies for a number of collinear spin configurations were mapped onto a classical Heisenberg model. The validity of the computational results was checked by calculations within generalized gradient approximation (GGA)\cite{pbe} and by choosing $U_d$ values of 4 and 6~eV. Similar to Cu$^{2+}$ oxides,\cite{tsirlin2010} the exchange-correlation potential (LDA $vs.$ GGA) has marginal effect on the spin model, whereas a change in $U_d$ only shifts the exchange couplings in a systematic way and keeps their ratios nearly constant. The DFT-based spin model was further studied by classical Monte-Carlo simulations with the \texttt{spinmc} algorithm of the ALPS package.\cite{alb2007}
\begin{table*}
\begin{minipage}{14cm}
\caption{\label{characters}
Characters of the irreducible representations of the little group of the propagation vector $\mathbf k=[0,\beta,0]$ for the space group $Pbcm$ ($a=e^{i\pi\beta}$) and the corresponding magnetic superspace groups. The decomposition of the magnetic representation $\Gamma_{\text{mag}}$ for all four-fold magnetic sites is: $\Gamma_{\text{mag}} = 3\Gamma_1+3\Gamma_2+3\Gamma_3+3\Gamma_4$.}
\begin{ruledtabular}
\begin{tabular}{cccccc}
	& $(E\,|\,0\,0\,0)$	& $(m_x\,|\,0\,\frac12\,0)$ & $(2_y\,|\,0\,\frac12\,0)$ & $(m_z\,|\,0\,0\,0)$ & Superspace group \\\hline
  $\Gamma_1$ & 1 & $a$ & $a$ & 1 & $Pbcm1'(0\beta0)000s$ \\
  $\Gamma_2$ & 1 & $-a$ & $a$ & $-1$ & $Pbcm1'(0\beta0)s0ss$ \\
  $\Gamma_3$ & 1 & $-a$ & $-a$ & 1 & $Pbcm1'(0\beta0)s00s$ \\
  $\Gamma_4$ & 1 & $a$ & $-a$ & $-1$ & $Pbcm1'(0\beta0)00ss$ \\
\end{tabular}
\end{ruledtabular}
\end{minipage}
\end{table*}
\begin{table*}
\caption{\label{operators}
Symmetry operators for the $Pbcm1'(0\beta0)s00s$ and $Pbcm1'(0\beta0)00ss$ magnetic superspace groups for origin shift $\delta=0,\frac14$ along $x_4$. The operators resulting from combination with the $(1'\,|\,0\,0\,0\,\frac12)$ generator, always present, are not shown ($-m$ means a ``time inversion'' operation, while $m$ is an operation without time inversion). The operators forming the common subgroups are printed in boldface.}
\begin{ruledtabular}
\begin{tabular}{rccc}
	& $Pbcm1'(0\beta0)s00s$, & $Pbcm1'(0\beta0)00ss$, & $Pbcm1'(0\beta0)00ss$,\\
	& origin at 0,0,0,0 & origin at 0,0,0,0 & origin at $0,0,0,\frac14$ \\\hline
  $E$ & $x_1,x_2,x_3,x_4,m$ & $\mathbf{x_1,x_2,x_3,x_4,m}$	& $\mathbf{x_1,x_2,x_3,x_4,m}$ \\
  $2_{1,z}$ & $-x_1,-x_2,x_3+\frac12,-x_4,m$	& $-x_1,-x_2,x_3+\frac12,-x_4+\frac12,m$ & $\mathbf{-x_1,-x_2,x_3+\frac12,-x_4,m}$ \\
  $2_{1,y}$ & $-x_1,x_2+\frac12,-x_3+\frac12,x_4+\frac12,m$ & $\mathbf{-x_1,x_2+\frac12,-x_3+\frac12,x_4+\frac12,m}$ & $\mathbf{-x_1,x_2+\frac12,-x_3+\frac12,x_4+\frac12,m}$ \\
  $2_x$ & $x_1,-x_2+\frac12,-x_3,-x_4+\frac12,m$ & $x_1,-x_2+\frac12,-x_3,-x_4,m$ & $\mathbf{x_1,-x_2+\frac12,-x_3,-x_4+\frac12,m}$ \\
  $\bar{1}$ & $-x_1,-x_2,-x_3,-x_4,m$ & $\mathbf{-x_1,-x_2,-x_3,-x_4,m}$ & $-x_1,-x_2,-x_3,-x_4+\frac12,m$ \\
  $m$ & $x_1,x_2,-x_3+\frac12,x_4,m$ & $x_1,x_2,-x_3+\frac12,x_4+\frac12,m$ & $x_1,x_2,-x_3+\frac12,x_4+\frac12,m$ \\
  $c$ & $x_1,-x_2+\frac12,x_3+\frac12,-x_4+\frac12,m$ & $\mathbf{x_1,-x_2+\frac12,x_3+\frac12,-x_4+\frac12,m}$ & $x_1,-x_2+\frac12,x_3+\frac12,-x_4,m$ \\\medskip
  $b$ & $-x_1,x_2+\frac12,x_3,x_4+\frac12,m$ & $-x_1,x_2+\frac12,x_3,x_4,m$ & $-x_1,x_2+\frac12,x_3,x_4,m$ \\
  & Subgroup & $P12_1/c11'(0\beta0)s0s$,  & $P22_12_11'(0\beta0)0s0s$,  
  \cr & & origin at $0,0,0,0$ & origin at $0,\frac14,0,0$ \\
\end{tabular}
\end{ruledtabular}
\end{table*}

\section{Results}
\label{results}
\subsection{Magnetic structure}
At room temperature (RT), BiMnFe$_2$O$_6$ crystallizes in an orthorhombic unit cell with $a=5.03590(3)$~\r A, $b=7.07342(4)$~\r A, $c=12.65425(6)$~\r A, and $Pbcm$ space symmetry.\cite{foot11} Below $T_N\simeq 220$~K, extra reflections appear on the NPD patterns (Fig.~\ref{NPD}). These reflections are of magnetic origin and can not be attributed to a structural phase transition because the x-ray diffraction experiment does not show any change down to $T=120$~K.\cite{yang2010} The magnetic reflections on the $T=7$~K pattern are indexed with a propagation vector $\mathbf k=[0,\beta,0]$ with $\beta=0.1379(1)$. The propagation vector is inside of the Brillouin zone and has a star with two arms {$\mathbf k,-\mathbf k$}. The little group of the propagation vector $G_k$ is $Pb2_1m$. In BiMnFe$_2$O$_6$, there are two symmetrically independent magnetic species in the nuclear $Pbcm$ structure: FeMn1 ($8e$: $0.4891,-0.1597,-0.6036$) and FeMn2 ($4a$: $0,0,\frac12$) (Fig.~\ref{structure}). In the little group $Pb2_1m$, this corresponds to two four-fold magnetic sites for the FeMn1 position and one four-fold magnetic site for the FeMn2 position (origin at $0,0,\frac14$). There are four one-dimensional irreducible representations (irreps) for the propagation vector $\mathbf k=[0,\beta,0]$ in the space group $Pbcm$; their characters are given in Table~\ref{characters}. The magnetic structure is transformed according to one of the irreps or their combination. Alternatively, we can describe the symmetry of the incommensurately modulated magnetic structure by embedding it into a higher-dimensional space and applying magnetic superspace groups defined in (3+1)-dimensional superspace.\cite{janner1980,schobinger2006,schobinger2007,petricek2010} The magnetic moment on atom $i$ is expressed as a vector function:
$$\mathbf M_i(x_4)=\mathbf M_{i0}+\sum_{n=1}^{N}[\mathbf M_{ins}\sin(2\pi nx_4)+\mathbf M_{inc}\cos(2\pi nx_4)],$$
where $n$ denotes terms of the Fourier series, \mbox{$x_4=\mathbf k(\mathbf T+\mathbf r_i)$} is an internal coordinate, $\mathbf T$ is the lattice translation of the nuclear structure, and $\mathbf r_i$ is the position of the atom $i$ in the unit cell of the nuclear structure. Monitoring of the intensity of the magnetic and nuclear reflections upon varying temperature revealed that there is no magnetic impact into the intensity of the nuclear reflections, and therefore $\mathbf M_{i0}=0$ for all magnetic sites. Since only the first-order satellites were observed, the Fourier series are reduced to the $n=1$ terms. A magnetic (Shubnikov) superspace group describing the transformations of the magnetic modulation waves can be set in correspondence with each irrep. The magnetic superspace groups are based on generators of the little group $G_k$, but the symmetry elements transforming the propagation vector $\mathbf k$ to $-\mathbf k$ also should be taken into account. This yields four possible magnetic superspace groups listed in Table~\ref{characters}. The explanation of the magnetic superspace group symbols is provided in Ref.~\onlinecite{petricek2010} 

All four magnetic superspace groups were tested in the refinement. Acceptable solutions were found in magnetic superspace groups $Pbcm1'(0\beta0)s00s$ and $Pbcm1'(0\beta0)00ss$, both with the same reliability factor for magnetic reflections $R_I^{\text{mag}}=0.057$. In spite of the relatively low reliability factor, the correspondence between the experimental and calculated NPD profile was far from ideal for both models. This indicates that the actual solution requires a combination of the irreps $\Gamma_3$ and $\Gamma_4$, and can be realized in a common subgroup of $Pbcm1'(0\beta0)s00s$ and $Pbcm1'(0\beta0)00ss$. This subgroup depends on a relative shift $\delta$ along the internal space of the conventional origins of two superspace groups. The list of operators of $Pbcm1'(0\beta0)s00s$ and their intersections with the operators of $Pbcm1'(0\beta0)00ss$ for \mbox{$\delta=0,\frac14$} are provided in Table~\ref{operators}. The resulting common subgroups are $P12_1/c11'(0\beta0)s0s$ and $P22_12_11'(0\beta0)0s0s$. For any $\delta$ values inequivalent to the cases mentioned above the common subgroup is $P12_111'(0\beta0)ss$.
 
The solutions in the $P12_1/c11'(0\beta0)s0s$ and $P22_12_11'(0\beta0)0s0s$ groups provide the same quality of the Rietveld fit and can not be distinguished on this basis. The $P12_1/c11'(0\beta0)s0s$ magnetic superspace symmetry results in a collinear magnetic structure with an antiferromagnetic transverse amplitude modulated wave with the magnetic moments confined to the $ac$ plane. The amplitude of the magnetic moment modulation varies from almost zero to 5.6~$\mu_B$, which is unrealistically high for the Mn$^{3+}$ and Fe$^{3+}$ cations. Thus, the $P12_1/c11'(0\beta0)s0s$ solution was ruled out. In the $P22_12_11'(0\beta0)0s0s$ model, there are three magnetic symmetrically unequivalent atoms: FeMn1a, FeMn1b, and FeMn2 (see Table \ref{coordinates}), all at the general four-fold sites of the $P22_12_1$ space group. The $(1'\,|\,0\,0\,0\,\frac12)$ operator of the magnetic superspace group requires the $\mathbf M_{i0}$ term to be zero and constrains the magnetic moment modulation functions to odd harmonics only, thereby resulting in the absence of the magnetic impact into intensity of the nuclear reflections and the absence of even-order magnetic satellites.\cite{petricek2010} The components of the magnetic moment modulation function for the general four-fold site are related by symmetry elements of the $P22_12_11'(0\beta0)0s0s$ magnetic superspace group, as shown in Table~\ref{functions}.

\begin{table*}
\caption{\label{coordinates}
Crystallographic parameters and atomic coordinates in BiMnFe$_2$O$_6$ at $T=7$~K and room temperature (RT). Atomic displacement parameters $U_{\iso}$ are given in $10^{-2}$~\r A$^2$. The symmetry operators of the $P22_12_1$ space group are listed in Table~\ref{operators}.}
\begin{ruledtabular}
\begin{tabular}{c@{\hspace{3em}}rrrr@{\hspace{3em}}rrrr}
	& \multicolumn{4}{c}{$T=7$~K} & \multicolumn{4}{c}{RT} \\\hline
 Space group & \multicolumn{4}{c}{$P22_12_1$} & \multicolumn{4}{c}{$Pbcm$}     \\
 $a$ (\r A)  & \multicolumn{4}{c}{5.02305(7)} & \multicolumn{4}{c}{5.03589(3)} \\
 $b$ (\r A)  & \multicolumn{4}{c}{7.06232(9)} & \multicolumn{4}{c}{7.07341(4)} \\
 $c$ (\r A)  & \multicolumn{4}{c}{12.6424(2)} & \multicolumn{4}{c}{12.65425(6)} \\\hline
 Atom & \multicolumn{1}{c}{$x$} & \multicolumn{1}{c}{$y$} & \multicolumn{1}{c}{$z$} & $U_{\iso}$ & \multicolumn{1}{c}{$x$} & \multicolumn{1}{c}{$y$} & \multicolumn{1}{c}{$z$} & \multicolumn{1}{c}{$U_{\iso}$} \\\hline
 Bi & 0.9711(3) & $-0.1301(2)$ & 0.751(1) & 0.27(4) & 0.9705(1) & $-0.1305(1)$ & $\frac34$ & 0.68(2) \\
 FeMn1a\footnote{FeMn1a and FeMn1b: 0.7Fe+0.3Mn;\quad FeMn2: 0.6Fe+0.4Mn} & 0.486(3) & $-0.160(2)$ & $-0.607(1)$ & 0.32(8) & 0.4891(3) & $-0.1597(2)$ & $-0.60355(7)$ & 0.80(3) \\ 
 FeMn1b$^\text{a}$ & $-0.488(3)$ & 0.160(3) & 0.607(1) & 0.32(8) & \\ 
 FeMn2$^\text{a}$ & 0.000(6) & 0.000(6) & 0.500(2) & 0.32(8) & 0 & 0 & $\frac12$ & 0.37(4) \\ 
 O1a & 0.157(3) & 0.439(3) & $-0.6349(9)$ & 0.47(2) & 0.1629(4) & 0.4372(3) & $-0.6366(2)$ & 0.65(5) \\
 O1b & $-0.162(3)$ & $-0.441(3)$ & 0.6386(9) & 0.47(2) & \\
 O2a & 0.672(2) & $-0.408(2)$ & $-0.5783(9)$ & 0.47(2) & 0.6641(4) & $-0.4077(4)$ & $-0.5765(2)$ & 0.96(5) \\
 O2b & $-0.660(2)$ & $-0.593(2)$ & 0.5746(9) & 0.47(2) & \\
 O3a & 0.787(3) & $\frac14$ & $\frac12$ & 0.47(2) & 0.7913(6) & $\frac14$ & $\frac12$ & 0.86(7) \\
 O3b & $-0.799(3)$ & $\frac34$ & $\frac12$ & 0.47(2) & \\
 O4 & 0.3444(6) & $-0.3044(4)$ & 0.753(1) & 0.47(2) & 0.3457(7) & $-0.3022(4)$ & $\frac34$ & 0.69(7) \\
\end{tabular}
\end{ruledtabular}
\end{table*}

\begin{figure}
\includegraphics{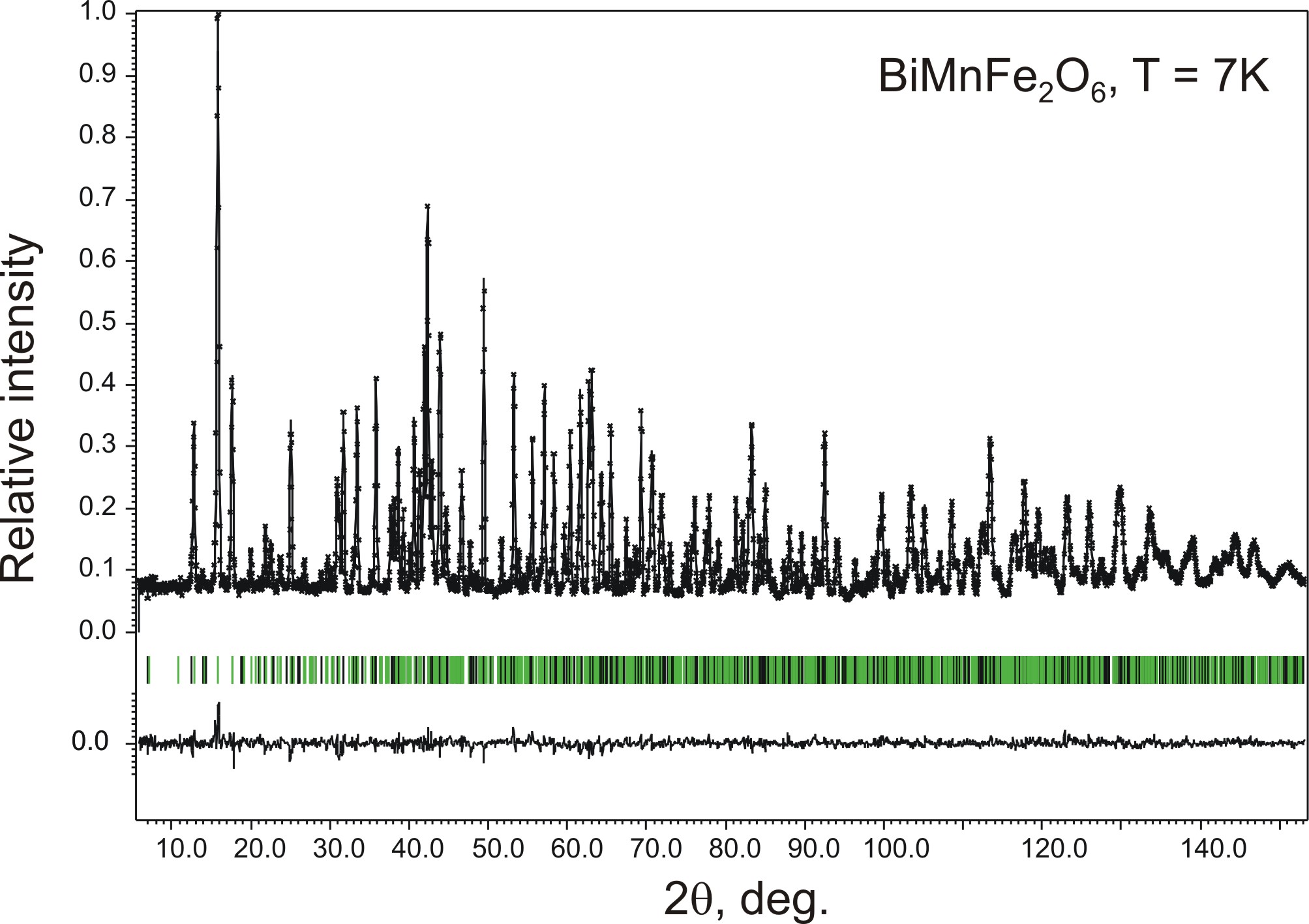}
\caption{\label{Rietveld}
(Color online) Experimental, calculated, and difference NPD pattern for BiMnFe$_2$O$_6$ at $T=7$~K. The black (dark) and green (light) bars mark the reflection positions for the nuclear and magnetic structures, respectively.
}
\end{figure}

\begin{figure}
\includegraphics{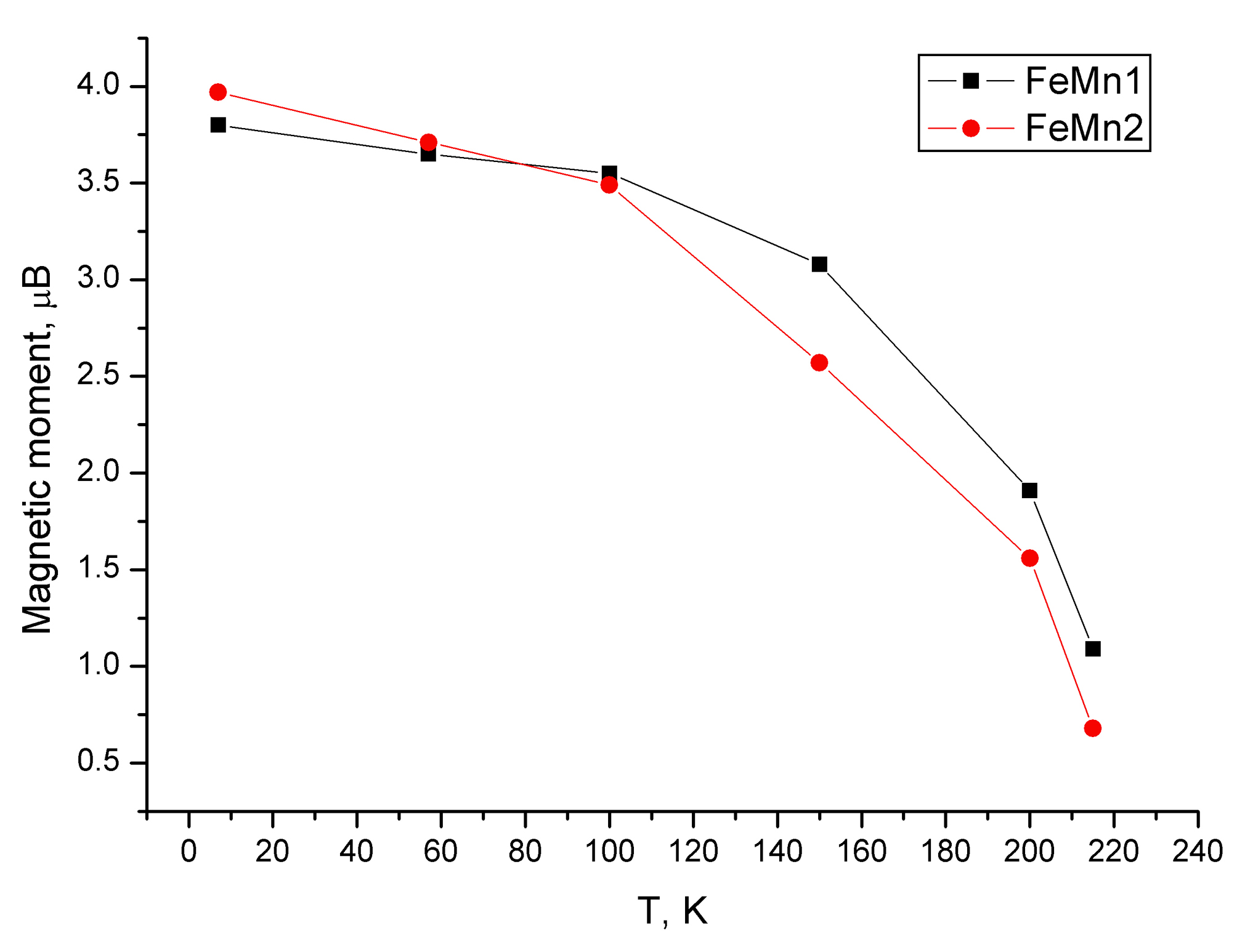}
\caption{\label{momtemp}
(Color online) Temperature dependence of the magnetic moment for the FeMn1 and FeMn2 positions.
}
\end{figure}

Since in the nuclear structure the atoms FeMn1a and FeMn1b are crystallographically equivalent, it is reasonable to assume that the modulation of the magnetic moment at these positions follows the same type of modulation waves. The solution was found with the following restrictions (not imposed by the magnetic superspace group) on the coefficients of the magnetic moment modulation functions: 
\begin{widetext}
\begin{center}
\begin{minipage}{12cm}
\begin{align*}
  M_{s,z}\text{(FeMn1a)}&=& M_{c,x}\text{(FeMn1a)}&=& M_{s,z}\text{(FeMn1b)}&=& M_{c,x}\text{(FeMn1b)}&=& M_1 \\
  M_{s,x}\text{(FeMn1a)}&=& -M_{c,z}\text{(FeMn1a)}&=& -M_{s,x}\text{(FeMn1b)}&=& M_{c,z}\text{(FeMn1b)}&=& M_2 \\
  M_{s,z}\text{(FeMn2)}&=& M_{c,x}\text{(FeMn2)} && && &=& M_3 \\
  M_{c,z}\text{(FeMn2)}&=& M_{s,x}\text{(FeMn2)} && && &=& 0 \\
\end{align*}
\end{minipage}
\end{center}
\end{widetext}
The refined magnetic moment components along the $b$ axis were smaller than their standard deviations and were fixed to zero. The refined magnetic parameters at different temperatures are provided in Table~\ref{moments}. The experimental, calculated, and difference NPD profiles at $T=7$~K are shown in Fig.~\ref{Rietveld}. The temperature dependencies of the ordered magnetic moment for the FeMn1 and FeMn2 positions are given in Fig.~\ref{momtemp}. Extrapolations of these dependencies with the $M=M_0[(1-T/T_N)^{\alpha}]^{\beta}$  function give coinciding $T_N$ values of 221(2)~K and 219(4)~K for the FeMn1 and FeMn2 positions, respectively, and the magnetic moments of $M_0\text{(FeMn1)}=3.76(4)$~$\mu_B$ and $M_0\text{(FeMn2)}=4.0(1)$~$\mu_B$. 

\begin{table*}
\begin{minipage}{14cm}
\caption{\label{functions}
Symmetry-imposed relations between the components of the magnetic moment modulation functions for the general four-fold site of the $P22_12_11'(0\beta0)0s0s$ magnetic superspace group with the origin at 0,$\frac14$,0,0.}
\begin{ruledtabular}
\begin{tabular}{cccc}
 Operator	& \multicolumn{3}{c}{Components of $\mathbf M_i(x_4)$} \\\hline
 $(E\,|\,0\,0\,0,0)$ & $M_x(x_4)$ & $M_y(x_4)$ & $M_z(x_4)$ \\
 $(2_x\,|\,0\,\frac12\,0, \frac12)$ & $M_x(-x_4+\frac12)$ & $M_y(-x_4)$ & $M_z(-x_4)$ \\
 $(2_y\,|\,0\,\frac12\,\frac12, \frac12)$ & $M_x(x_4)$ & $M_y(x_4+\frac12)$ & $M_z(x_4)$ \\
 $(2_z\,|\,0\,0\,\frac12, 0)$ &	$M_x(-x_4+\frac12)$	& $M_y(-x_4+\frac12)$ & $M_z(-x_4)$ \\
\end{tabular}
\end{ruledtabular}
\end{minipage}
\end{table*}

\begin{table*}
\begin{minipage}{14cm}
\caption{\label{moments}
Refined magnetic parameters for BiMnFe$_2$O$_6$ at different temperatures (see text for notations). Reliability factors are listed in the order of $R_I$ (overall), $R_I$ (nuclear reflections), $R_I$ (magnetic satellites), $R_P$.}
\begin{ruledtabular}
\begin{tabular}{ccccccc}
	$T$ &	$M_1$     &	$M_2$	    & $M_3=M$(FeMn2)  & $M$(FeMn1) & $\beta$ & $R$-factors \\
	(K) & ($\mu_B$) & ($\mu_B$) & ($\mu_B$)       & ($\mu_B$)  &         &             \\\hline
  7 & 3.25(2) & 1.94(2) & 3.98(3) & 3.80(3) & 0.13801(8) & 0.017, 0.014, 0.028, 0.036 \\
 57 & 3.13(2) & 1.87(2) & 3.71(2) & 3.65(3) & 0.13740(4) & 0.016, 0.013, 0.025, 0.030 \\
100 & 3.09(4) & 1.75(3) & 3.50(4) & 3.55(5) & 0.1316(1) & 0.025, 0.022, 0.040, 0.048 \\
150	& 2.74(6) & 1.40(5) & 2.57(6) & 3.08(8) & 0.1232(2) & 0.030, 0.029, 0.036, 0.067 \\
200	& 1.69(6) & 0.88(5)	& 1.56(6)	& 1.91(8)	& 0.1175(3)	& 0.029, 0.027, 0.044, 0.051 \\
215	& 0.95(9)	& 0.54(9)	& 0.68(9)	& 1.1(1) & 0.1155(9) & 0.028, 0.028, 0.036, 0.049 \\
\end{tabular}
\end{ruledtabular}
\end{minipage}
\end{table*}

\begin{figure}
\includegraphics{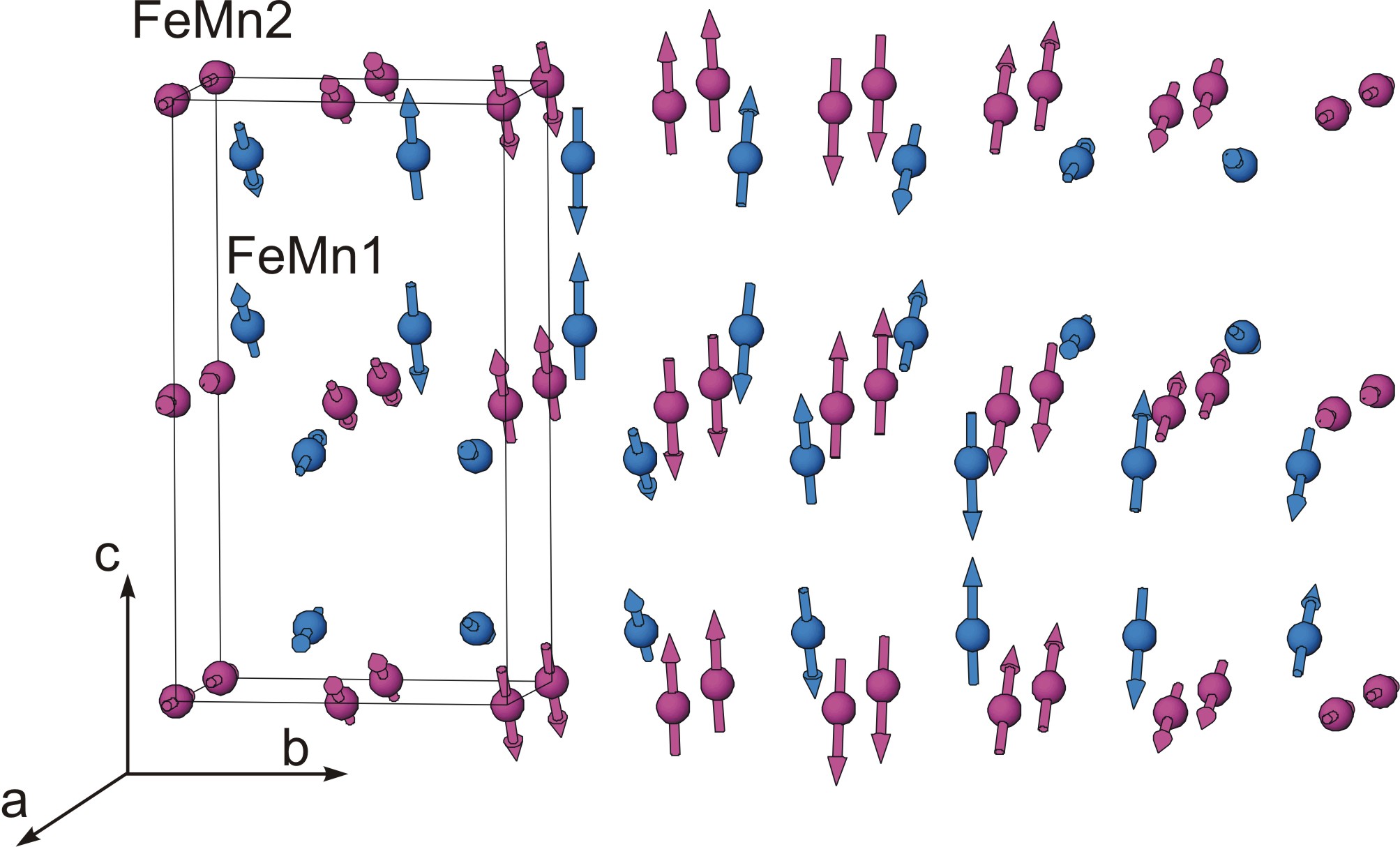}
\caption{\label{spinstr}
(Color online) The arrangement of magnetic moments in BiMnFe$_2$O$_6$. The unit cell for the nuclear structure is outlined.
}
\end{figure}

The spin arrangement in the BiMnFe$_2$O$_6$ magnetic structure is shown in Fig.~\ref{spinstr}. BiMnFe$_2$O$_6$ adopts a spiral magnetic structure consisting of antiferromagnetic helixes propagating along the $b$-axis with a period of $\,\simeq\!3.5b$. The helixes are associated with the FeMn1 and FeMn2 atomic chains running along the $b$-axis, where the magnetic atoms are separated by $b/2$. The magnetic moments rotate for $\pi(1+\beta)\simeq 204.8$~deg about the $b$-axis on going between the adjacent magnetic atoms in the chains. The refined parameters $M_1$, $M_2$, and $M_3$ represent the magnetic moment of the FeMn1 atom $\left[M\text{(FeMn1)}=\sqrt{M^2_1+M^2_2}\right]$, the magnetic moment of the FeMn2 atom $\left[ M\text{(FeMn2)}=|M_3|\,\right]$, and the phase shift $\varphi$ between the FeMn1 and FeMn2 helixes \mbox{($\tan\varphi=-{\frac{M_2}{M_1}}$)}. 

The refined structural parameters and interatomic distances at $T=7$~K (space group $P22_12_1$) and RT (space group $Pbcm$) are listed in Tables \ref{coordinates} and \ref{distances}, respectively. The crystal structure at $T=7$~K was refined with fixed parameters of the magnetic structure. The crystal structures at both temperatures are virtually identical indicating that the magnetic ordering does not influence the nuclear structure. With the $P22_12_11'(0\beta0)0s0s$ magnetic symmetry, the space group of the average nuclear structure should be $P22_12_1$ (point group 222). Although the spiral magnetic ordering eliminates the inversion center, it does not create a polar direction. Indeed, no indication of ferroelectricity below $T_N$ was found in BiMnFe$_2$O$_6$ by dielectric permittivity measurements.\cite{yang2010} 

\begin{table*}
\begin{minipage}{16cm}
\caption{\label{distances}
Selected interatomic distances (in~\r A) in BiMnFe$_2$O$_6$ at $T=7$~K and room temperature (RT).}
\begin{ruledtabular}
\begin{tabular}{cc@{\hspace{4em}}cc@{\hspace{4em}}cc}
 Bond & Length & Bond & Length & Bond & Length \\\hline
 $T=7$~K & & & & & \\					
 Bi--O1a & 2.20(2), 2.69(2) & FeMn1a--O1b & 1.95(2) & FeMn2--O1a & 1.93(3) \\
 Bi--O1b & 2.16(2), 2.70(2) & FeMn1a--O2a & 2.02(2), 2.56(2) & FeMn2--O1b & 1.98(3) \\
 Bi--O4 & 2.243(3) & FeMn1a-O2b & 1.97(2) & FeMn2--O2a & 2.03(3)\
 \cr & & FeMn1a--O3b & 2.07(2) & FeMn2--O2b & 2.06(3) \
 \cr & & FeMn1a--O4 & 2.00(2) & FeMn2--O3a & 2.07(4) \
 \cr & & FeMn1b--O1a & 1.95(2) & FeMn2--O3b & 2.03(4) \
 \cr & & FeMn1b--O2a & 1.99(2) & & \	
 \cr & & FeMn1b--O2b & 1.99(2), 2.51(2) & & \		
 \cr & & FeMn1b--O3a & 2.03(2) & & \\
 \cr & & FeMn1b--O4 & 1.92(2) & & \\\hline	
 RT & & & & & \\					
 Bi--O1($\times$2) & 2.207(2) & FeMn1--O1 & 1.928(3) & FeMn2--O1($\times$2) & 1.964(2) \\
 Bi--O1($\times$2) & 2.686(2) & FeMn1--O2 & 1.973(3) & FeMn2--O2($\times$2) & 2.056(2) \\
 Bi--O4 & 2.246(3) & FeMn1--O2 & 1.992(3) & FeMn2-O3($\times$2) & 2.057(2)\\
 Bi--O2($\times$2) & 2.696(2) & FeMn1--O4 & 2.007(2) & & \\		
 Bi--O4 & 2.816(3) & FeMn1--O3 & 2.029(3) & & \		
 \cr & & FeMn1--O2 & 2.489(2) & & \\		
\end{tabular}
\end{ruledtabular}
\end{minipage}
\end{table*}

\subsection{Electronic structure}
Owing to the complex crystal structure, an empirical assignment of individual exchange couplings in BiMnFe$_2$O$_6$ is a formidable challenge. The problem can be solved by electronic structure calculations that evaluate individual couplings and, therefore, establish a reliable microscopic magnetic model. Recent studies prove the remarkable accuracy of DFT for diverse correlated systems, including frustrated magnets with highly intricate spin lattices.\cite{gorelov2010,mazurenko2008,tsirlin2010a} Prior to computing exchange integrals, we discuss the electronic structure of BiMnFe$_2$O$_6$, which is a key to understanding the magnetic behavior. 

Computational analysis of BiMnFe$_2$O$_6$ is complicated by the intrinsic disorder of Fe and Mn atoms. Unfortunately, state-of-the-art computational tools, such as virtual crystal approximation (VCA) or coherent potential approximation (CPA), cannot be applied to this case, since they provide an averaged description of a disordered system, while we are targeting magnetic properties that depend on local interactions. The insulating nature of BiMnFe$_2$O$_6$ \cite{yang2010} suggests localized 3$d$ electrons of transition metals; therefore, the distinct spin-$\frac52$ Fe$^{3+}$ ($d^{5}$) and \mbox{spin-2} Mn$^{3+}$ ($d^{4}$) sites are randomly distributed on the spin lattice. Further, orbital ordering for Mn$^{3+}$ results in dramatic differences among the Fe--O--Fe, Mn--O--Mn, and Fe--O--Mn superexchanges (see below). An averaged VCA or CPA picture would not reproduce any of these features. To capture effects arising from the localized moments of Fe$^{3+}$ and Mn$^{3+}$, we calculated exchange integrals for several systems with ordered Fe and Mn atoms, and accessed all possible scenarios of the superexchange. In these calculations, we kept the experimental atomic positions, but imposed different arrangements of Fe and Mn, spanning purely Fe (BiFe$_3$O$_6$) and purely Mn (BiMn$_3$O$_6$) cases as well as intermediate (BiMnFe$_2$O$_6$ and BiMn$_2$FeO$_6$) configurations.

\begin{figure}
\includegraphics{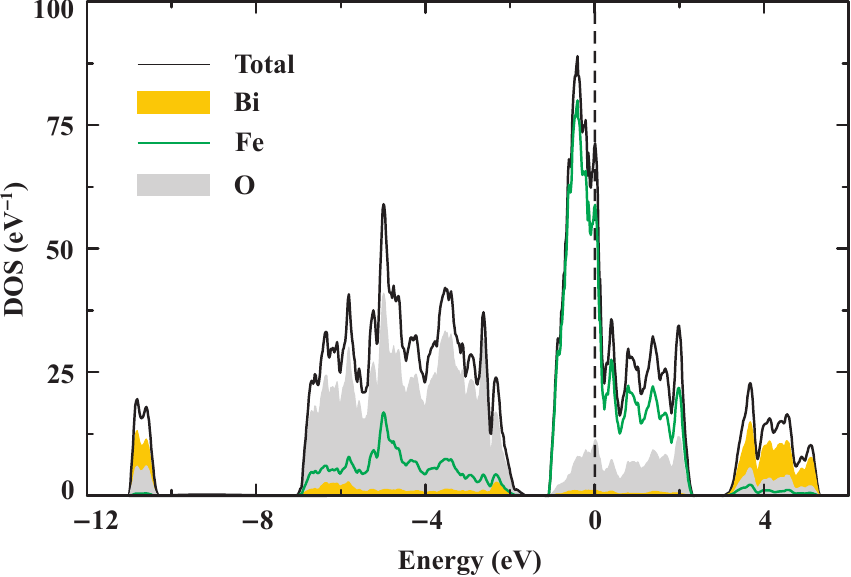}
\caption{\label{ldados}
(Color online) LDA density of states for BiFe$_3$O$_6$. The Fermi level is at zero energy. The gapless energy spectrum is caused by underestimation of the electronic correlations in LDA.}
\end{figure}

\begin{figure}
\includegraphics{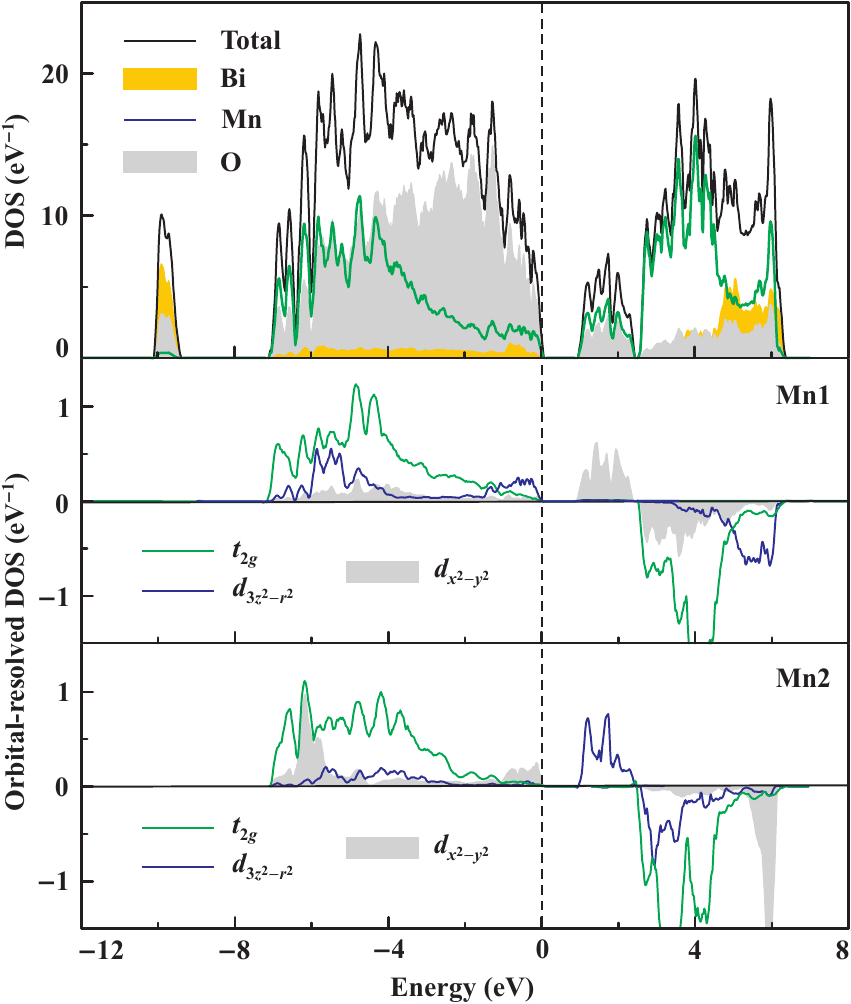}
\caption{\label{lda+udos}
(Color online) Top: LSDA+$U$ DOS for one spin channel in the ground-state AFM configuration of BiMn$_3$O$_6$ ($U_d$ = 5~eV). Middle and bottom: orbital-resolved DOS for 3$d$ states of Mn$^{3+}$. Four out of five $d$ orbitals show filled spin-up and empty spin-down states, whereas the states of the remaining $d$ orbital ($d_{x^{2}-y^{2}}$ for Mn1 and $d_{3z^{2}-r^{2}}$ for Mn2) are mostly above the Fermi level for both spin directions. The Fermi level is at zero energy.}
\end{figure}

The LDA band structure depends only marginally on the Fe/Mn ordering. The density of states (DOS) features Bi 6$s$ bands below $-10$~eV, O $2p$ states between $-7$~eV and $-2$~eV, transition-metal 3$d$ states at the Fermi level, and Bi $6p$ states above 3~eV (Fig.~\ref{ldados}). The effect of substituting Mn for Fe is a change in the electron count and the ensuing shift of the Fermi level within the $3d$ bands. Irrespective of the Fe/Mn ratio, LDA band structures are metallic due to the heavy underestimation of strong electronic correlations. The insulating spectrum is correctly reproduced by DFT+$U$ (see upper panel of Fig.~\ref{lda+udos}). In particular, the band gaps of about 1.5~eV for BiFe$_3$O$_6$ and 1.1~eV for BiMn$_3$O$_6$ at $U_d=5$~eV are consistent with the black color of the compound.

Superexchange couplings in insulating compounds intimately depend on the orbital state of the transition metal. Five unpaired electrons of Fe$^{3+}$ fill five $d$ orbitals and leave no orbital degrees of freedom. By contrast, Mn$^{3+}$ has four unpaired electrons only, and hence one of the $d$ orbitals is unoccupied in the Mott-insulating state. Since there is a sizable crystal-field splitting (about 1.5~eV) driven by the octahedral coordination of the FeMn1 and FeMn2 sites, the three $t_{2g}$ states are half-occupied, while the two $e_g$ states have to share the remaining electron. The half-filled orbital is picked out by a weak distortion of the octahedral local environment. In the following, we analyze such distortions in more detail to determine which of the two $e_g$ orbitals is half-filled in the insulating state.

The FeMn1 position reveals one long bond of about 2.5~\r A (FeMn1--O2, Table~\ref{distances}) that we choose as the local $z$ axis. According to simple electrostatic arguments of crystal-field theory, the long bond along $z$ shifts the $d_{3z^{2}-r^{2}}$ orbital down in energy with respect to the $d_{x^{2}-y^{2}}$ orbital. The opposite scenario is found for FeMn2, where the local environment is a squeezed octahedron. The short FeMn2--O1 bond (about 1.96~\r A, Table~\ref{distances}), which we take as the local $z$-axis, shifts the $d_{3z^{2}-r^{2}}$ orbital up in energy. Therefore, the unpaired electron of Mn$^{3+}$ should occupy the $d_{3z^{2}-r^{2}}$ orbital for FeMn1 and the $d_{x^{2}-y^{2}}$ orbital for FeMn2. This conclusion is verified by DFT+$U$ calculations that place unpaired electrons on the respective orbitals, while shifting both spin-up and spin-down states for the remaining $e_g$ orbital ($d_{x^{2}-y^{2}}$ and $d_{3z^{2}-r^{2}}$ for FeMn1 and FeMn2, respectively) above the Fermi level (see Fig.~\ref{lda+udos}).\cite{foot12}
 
The orbital order in BiMn$_3$O$_6$ is formally of the ferro-type within the FeMn1 and FeMn2 sublattices, yet of the antiferro-type between the two sublattices. However, this notation is deceptive because of the different local axes on neighboring atoms. The magnetic model largely deviates from the conventional scenario\cite{kugel1982} of AFM superexchange for ferro-type orbital order and ferromagnetic (FM) superexchange for antiferro-type orbital order. In the following, we show that the peculiar arrangement of empty $e_g$ orbitals in BiMn$_3$O$_6$ induces sizable FM superexchange for nearly all couplings and alters the spiral ground state that arises from the purely AFM spin lattice of BiFe$_3$O$_6$.

\subsection{Microscopic magnetic model}
Exchange couplings calculated for different superexchange scenarios (Fe--O--Fe, \mbox{Mn--O--Mn}, Fe--O--Mn, and Mn--O--Fe) are listed in Table \ref{couplings}. The Fe--O--Fe, Mn--O--Mn, and Fe--O--Mn superexchanges show sharp differences for most of the couplings, while the Fe--O--Mn and Mn--O--Fe cases are only different for interactions between the FeMn1 and FeMn2 sublattices (i.e., when the two metal sites are not related by symmetry). In Table~\ref{couplings}, we restrict ourselves to short-range couplings matching direct connections between the FeMn octahedra. Long-range interactions are expected to be weak, as confirmed by the following qualitative argument. Long-range superexchange requires suitable overlap of atomic orbitals along a M--O--O--M (or even more complex) pathway, and can be achieved for one orbital channel only. By contrast, short-range superexchange is possible for most of the orbital channels, and should therefore dominate in systems with several magnetic orbitals. To verify this conclusion for BiMnFe$_2$O$_6$, we calculated the exchange couplings within the crystallographic unit cell and within a supercell doubled along $a$. The resulting values of $J_i$ agreed within $10\,$\% and indicated weak long-range couplings in BiMnFe$_2$O$_6$. Below, we demonstrate that our minimum microscopic model, restricted to short-range couplings, is sufficient to explain the spiral magnetic structure of BiMnFe$_2$O$_6$ and the lack of ferroelectricity in this compound. 

\begin{table*}
\begin{minipage}{14cm}
\caption{\label{couplings}
Interatomic distances (in~\r A) and leading exchange couplings (in~K) in BiMnFe$_2$O$_6$ calculated with the supercell procedure (LSDA+$U$, $U_d$ = 5~eV) for different scenarios of superexchange. Negative $J_i$ denotes FM coupling. The intralayer couplings $J_1-J_7$ and $J_9$ are depicted in Fig.~\ref{spinlattice}, whereas $J_8$ connects the layers along $c$.}
\begin{ruledtabular}
\begin{tabular}{cc@{\hspace{3em}}ccc}
 & Distance & \multicolumn{3}{c}{Exchange couplings $J_i$} \\\hline
 & & Fe--O--Fe & Mn--O--Mn & Fe--O--Mn/Mn--O--Fe \\
 $J_1$ & 2.916 & 1 & $-1$ & $-50$ \\
 $J_2$ & 3.010 & 19 & $-9$ & 29/10 \\
 $J_3$	& 3.101 & 25 & $-38$ & 2/$-23$ \\
 $J_4$ & 3.462 & 26 & 7 & $-38$ \\
 $J_5$ & 3.537 (FeMn2) & 30 & 13 & 43 \\
 $J_6$ & 3.538 (FeMn1) & 61 & $-8$ & 48 \\
 $J_7$ & 3.685 & 37 & $-1$ & 23/30 \\
 $J_8$ & 3.706 & 74 & 79 & 127 \\
 $J_9$ & 3.759 & 66 & $-12$ & 20/70 \\
\end{tabular}
\end{ruledtabular}
\end{minipage}
\end{table*}

\begin{figure}
\includegraphics{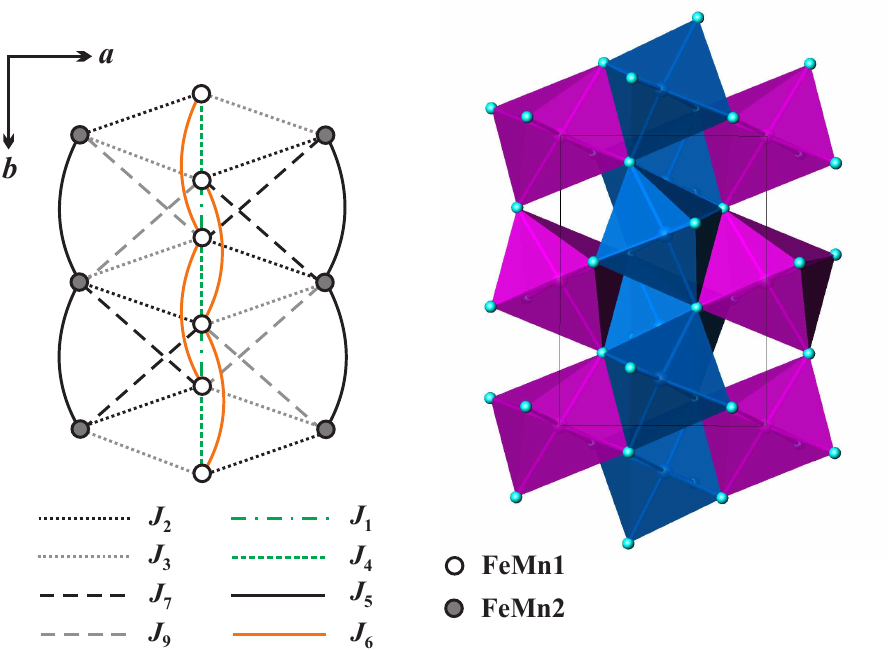}
\caption{\label{spinlattice}
(Color online) Spin lattice (left) and the respective part of the crystal structure (right) showing frustrated exchange couplings in the $ab$ plane. Open and filled circles denote the FeMn1 and FeMn2 positions, respectively.}
\end{figure}

The spin lattice of BiMnFe$_2$O$_6$ incorporates nine inequivalent exchanges (Fig.~\ref{spinlattice} and Table~\ref{couplings}). Eight of these couplings are found in the $ab$ plane, whereas $J_8$ connects the layers along~$c$. The interlayer coupling is weakly influenced by the Fe/Mn substitution, and remains one of the leading AFM interactions for all superexchange scenarios (Table \ref{couplings}). The robustness of $J_8$ should be traced back to the orbital order for Mn$^{3+}$. The local $z$-axis of the FeMn1 octahedra roughly matches the crystallographic $c$ direction. Therefore, the replacement of Fe by Mn results in an empty $d_{x^{2}-y^{2}}$ orbital (Fig.~\ref{lda+udos}) that contributes weakly to the superexchange along $c$. By contrast, most of the intralayer couplings are heavily affected by changing Mn for Fe.

In BiFe$_3$O$_6$, eight intralayer couplings are AFM. Excluding the apparently weak $J_1$, we arrive at seven AFM couplings ranging from 19 K to 74 K (Table~\ref{couplings}). Triangular loops abound (Fig.~\ref{spinlattice}) and lead to the strong frustration of the spin lattice. This frustration is largely released by Mn$^{3+}$ that renders most of the intralayer couplings FM, and reduces the remaining AFM couplings below 13 K (Table \ref{couplings}). The Fe--O--Mn and Mn--O--Fe cases are intermediate, with partially reduced AFM couplings. Large FM contributions to the superexchange can be traced back to empty $d$ orbitals of Mn$^{3+}$. These orbitals provide the strong $\sigma$-overlap with oxygen orbitals and a leading contribution to the superexchange.

To evaluate the ground state of the proposed model, we performed classical Monte-Carlo simulations for a \mbox{$16\times16\times16$} finite lattice with periodic boundary conditions. The unit cell of the spin lattice incorporated six magnetic atoms within one layer (half of the crystallographic unit cell). The temperature was set to 10 K, well below the ordering temperature $T_N$ (see further in this section). Spin-spin correlation functions for purely Fe$^{3+}$ ($S=\frac52$) and purely Mn$^{3+}$ ($S=2$) lattices are given in Table~\ref{correlations}. The correlation functions normalized for $S^{2}$ are $-1$ for antiparallel spins and $+1$ for parallel spins. Intermediate values indicate non-collinear configurations. 

\begin{table}
\caption{\label{correlations}
Normalized spin-spin correlations $\langle S_iS_j\rangle/S^2$ for BiFe$_3$O$_6$ (Fe$^{3+}$, $S=\frac52$) and BiMn$_3$O$_6$ (Mn$^{3+}$, $S=2$).}
\begin{minipage}{6cm}
\begin{ruledtabular}
\begin{tabular}{crr}
 & BiFe$_3$O$_6$ & BiMn$_3$O$_6$ \\\hline
 $J_1$ & 0.288 & 0.957 \\
 $J_2$ & 0.240 & 0.954 \\
 $J_3$ & 0.234 & 0.974 \\
 $J_4$ & $-0.835$ & 0.945 \\
 $J_5$ & $-0.693$ & 0.940 \\
 $J_6$ & $-0.706$ & 0.951 \\
 $J_7$ & $-0.827$ & 0.945 \\
 $J_8$ & $-0.992$ & $-0.993$ \\
 $J_9$ & $-0.835$ & 0.956 \\
\end{tabular}
\end{ruledtabular}
\end{minipage}
\end{table}

We consider the Fe$^{3+}$ case first. The normalized spin-spin correlation for $J_8$ is close to $-1$; therefore, the interlayer ordering is collinear and AFM. Similar correlations on the $J_5$ and $J_6$ bonds indicate the same propagation vector along $b$ for the FeMn1 and FeMn2 sublattices. Further on, similar correlations on the $J_2$ and $J_3$ as well as on $J_7$ and $J_9$ bonds signify the same magnetic order along the respective bonds and, consequently, a twice shorter periodicity along $b$ for the magnetic unit cell compared to the crystallographic unit cell. To find out the ordering pattern along $a$, we note that the respective FeMn2 atoms are connected via $J_2$ and $J_3$ or $J_7$ and $J_9$ bonds. The correlations along these paths are different; therefore, the spins on the FeMn2 atoms should be parallel (i.e., the magnetic moment rotates for a certain angle $\varphi_1$ on the $J_2$ bond and for the opposite angle $-\varphi_1$ on the $J_3$ bond). Thus, the propagation vector of the magnetic structure is $[0,\beta,0]$, in agreement with the experiment. The principal spin arrangement is described by three parameters, which are the angles between the spins on the $J_5$ ($J_6$), $J_2$, and $J_7$ bonds. According to our simulations, these angles are $\varphi=226$~deg, $\varphi_1=76$~deg, and $\varphi_2=146$~deg, respectively, in remarkable agreement with the experimental values of $\varphi=206$~deg, $\varphi_1=39$~deg, and $\varphi_2=167$~deg. Our microscopic model for BiFe$_3$O$_6$ reproduces the experimental magnetic structure of BiMnFe$_2$O$_6$ quite well. The remaining discrepancies are likely related to the partial replacement of Fe by Mn.

The complete substitution of Mn for Fe changes the magnetic ground state. According to Table \ref{correlations}, most of the normalized spin-spin correlations in BiMn$_3$O$_6$ are close to +1. The only negative correlation refers to the $J_8$ bond and indicates AFM interlayer coupling. The intralayer ordering is now collinear FM due to predominantly FM exchange couplings ($J_1$, $J_2$, $J_3$, $J_6$, $J_7$, and $J_9$, see Table \ref{couplings}). The spin lattice is still frustrated by AFM couplings $J_4$ and $J_5$, which, however, are not strong enough to induce the spiral order. While BiFe$_3$O$_6$ and BiMn$_3$O$_6$ present two opposite scenarios of strong and weak frustration, respectively, BiMnFe$_2$O$_6$ lies between these distinct regimes. The spiral ground state of this compound is driven by the frustration of intralayer exchange couplings that remain predominantly AFM for mixed Fe--O--Mn superexchange pathways (see Table \ref{couplings}). 

For an additional test of our microscopic model, we calculated magnetic susceptibility ($\chi$) and estimated the N\'eel temperature $T_N$ as the position of the kink in the temperature dependence of $\chi$. Transition temperatures for BiFe$_3$O$_6$ (420~K) and BiMn$_3$O$_6$ (150~K) reasonably agree with the experimental $T_N$ of 220~K that lies between the two calculated values.

\section{Discussion and summary}
\label{discussion}
Our experimental and computational study provides microscopic insight into the physics of BiMnFe$_2$O$_6$. This compound features a strongly frustrated spin lattice with predominantly AFM exchange couplings that induce the spiral magnetic order. The comparison between BiFe$_3$O$_6$ and BiMn$_3$O$_6$ suggests that AFM exchange couplings are an essential prerequisite for the spiral magnetic order. The crystal structure itself may still allow for different ground states; therefore, it is the transition-metal cation that determines the type of the long-range magnetic order. In this respect, first reports on cation substitution\cite{yang2010} look promising, because the change in the Fe/Mn ratio or an incorporation of other magnetic cations could alter the magnetic structure and other physical properties. 

BiMnFe$_2$O$_6$ is a magnetic compound that conforms to two mechanisms of ferroelectricity. First, lone pairs of Bi$^{3+}$ induce polar displacements. Second, the non-collinear magnetic structure breaks the inversion symmetry and allows for magnetic-field-induced electric polarization. However, none of the two mechanisms succeed in rendering the compound ferroelectric. Bi$^{3+}$-related polar displacements form an antiferroelectric pattern that leads to zero net polarization. The electronic mechanism meets a similar obstacle of the AFM interlayer coupling and the ensuing non-polar (albeit non-centrosymmetric) magnetic structure. The lack of polarity in both atomic and magnetic structures naturally explains the absence of the ferroelectric response in BiMnFe$_2$O$_6$ below $T_N$.\cite{yang2010} 

Our results demonstrate that the simple criteria of ferroelectricity and multiferroicity are not universal, since the polarization induced by any kind of polar distortion (atomic displacement or spin spiral) can be wiped out by an overall antiferroelectric/antiferromagnetic order. The combination of lone-pair and transition-metal cations does not necessarily lead to a magnetic ferroelectric, whereas an arbitrary incommensurate magnetic structure may not allow for the electronic mechanism of ferroelectricity. These simple observations make the search for multiferroics a challenge, and put forward the charge-ordering mechanism as a more robust approach to the design of magnetoelectric materials.

The cancellation of polarity in an incommensurate magnetic structure has been proposed for the spin-chain cuprate NaCu$_2$O$_2$\cite{capogna2010} and for a number of layered compounds, such as $\alpha$-CaCr$_2$O$_4$\cite{chapon2011,toth2011} and \mbox{$\alpha$-SrCr$_2$O$_4$},\cite{dutton2011} that were considered as potential multiferroics. The non-frustrated AFM interlayer coupling seems to be a general obstacle for ferroelectricity in layered systems, like $\alpha$-CaCr$_2$O$_4$,\cite{singh} or in three-dimensional systems with two-dimensional frustrated units, as in BiMnFe$_2$O$_6$. To overcome this problem, one has to design materials with FM or frustrated interlayer couplings. In the case of BiMnFe$_2$O$_6$, cation substitutions preserve the long-range order \cite{yang2010} and can be promising for tuning this compound toward ferroelectric and possibly multiferroic behavior.

In summary, we have solved the spiral magnetic structure of BiMnFe$_2$O$_6$ and proposed a microscopic magnetic model for this compound. The ground state features spins lying in the $ac$ plane and propagating along $b$ by a $\,\simeq\!204.8$ deg rotation about the $b$ axis. The two inequivalent FeMn positions reveal the same propagation vector. The spiral magnetic structure is driven by the strong frustration of antiferromagnetic exchange couplings on a complex spin lattice in the $ab$ plane. However, the coupling along the $c$ direction is non-frustrated, leading to antiparallel spin arrangement in neighboring layers. The resulting magnetic structure is non-polar and, alike the centrosymmetric (antiferroelectric) atomic structure, precludes the ferroelectric behavior of BiMnFe$_2$O$_6$.

\acknowledgments
We are grateful to Oleg Janson for granting access to the \texttt{flyswatter} program and fruitful discussions. A.T. was funded by Alexander von Humboldt Foundation. Contributions by Dr. Judith Stalick of NIST are gratefully acknowledged. Use of the National Synchrotron Light Source, Brookhaven National Laboratory, was supported by the U.S. Department of Energy, Office of Basic Energy Sciences, under Contract No. DE-AC02-98CH10886.


\begin{thebibliography}{42}
\bibitem [{foo({\natexlab{a}})}]{foot11}%
  \BibitemOpen
  \href@noop {} {} \bibinfo {note} {The error bars for unit
  cell parameters are derived from the structure refinement.}\BibitemShut
  {Stop}%
\bibitem [{foo({\natexlab{b}})}]{foot12}%
  \BibitemOpen
  \href@noop {} {} \bibinfo {note} {The opposite orbital
  configuration with the half-filled $d_{x^{2}-y^{2}}$ orbital for FeMn1 and
  $d_{3z^{2}-r^{2}}$ orbital for FeMn2 could not be stabilized.}\BibitemShut
  {Stop}%
\end{thebibliography}
%

\end{document}